\documentclass[10pt,osajnl,superscriptaddress,aps,twocolumn]{revtex4-1}
\usepackage[latin1]{inputenc}
\usepackage{amsmath}
\usepackage{mathptmx}

\begin{document}

\title{A Scalar Wigner Theory for Polarized Light in Nonlinear Kerr Media}

\author{T. Hansson} 
\affiliation{Department of Physics, Ume{\aa} University, SE-901~87 Ume{\aa}, Sweden }

\author{E. Wallin} 
\affiliation{Department of Physics, Ume{\aa} University, SE-901~87 Ume{\aa}, Sweden }

\author{G. Brodin} 
\affiliation{Department of Physics, Ume{\aa} University, SE-901~87 Ume{\aa}, Sweden }

\author{M. Marklund}
\email[E-mail address:]{mattias.marklund@physics.umu.se}
\altaffiliation[Also at: ]{Applied Physics, Chalmers University of Technology, SE--412 96 G\"oteborg, Sweden}
\affiliation{Department of Physics, Ume{\aa} University, SE-901~87 Ume{\aa}, Sweden }

\begin{abstract}
A scalar Wigner distribution function for describing polarized light is proposed in analogy with the treatment of spin variables in quantum kinetic theory. The formalism is applied to the propagation of circularly polarized light in nonlinear Kerr media and an extended phase space evolution equation is derived along with invariant quantities. We further consider modulation instability as well as the extension to partially coherent fields.
\end{abstract}

\maketitle

\section{Introduction}

The field of statistical optics, and its corresponding quantum statistical version, is a field where basic science questions meet applications, the latter in the form of, e.g., optical communication systems. Formalisms developed for the quantum realm has been transferred to classical problems, and the strong coupling to kinetic descriptions of plasmas and quantum kinetics of particles has given rise to interesting crossbreeding between seemingly disparate areas of research \cite{Hillery-etal, Lee, Schleich}. The description of atomic, molecular and optical physics, in particular quantum optical systems, in terms of statistical quantities lends itself to a multitude of applications, due to its direct relevance for the interpretation of experimental results (see, e.g., \cite{Bertet-etal}). The goal here is less ambitious; we are interested in the possibility of formulating and interpreting a scalar (quasi-)distribution function of polarized light. The extension of our results to applications will be left for future studies (something we discuss further in the Conclusions).

The spin and orbital angular momentum of light has over the last few years attracted an increasing interest, for a number of reasons. One such aspect lies in the possibility to extract more information from light sources, taking the angular momentum of light into account \cite{Information}. There has also been interest in this field due to the possibility to manipulate material systems through optical tweezers centered around tailored light sources with angular momentum \cite{Tweezers}. Thus, the dynamics of the polarization state of light, and its interaction with matter, lies at the heart of some recent applications.
Moreover, the spin degrees-of-freedom is something that recently has attracted interest in the field of quantum plasmas \cite{Eliasson-Shukla, Marklund-Brodin, Brodin-Marklund, Mahajan-Asenjo}, much due to the fact that it shows a potential to lead to new and important applications (see, e.g., \cite{Meta, Soliton, Asenjo-etal,Lundin}). We will here adopt an approach common to some of these works, in particular the approach of obtaining a scalar  distribution function from vector degrees-of-freedom using a suitable projection operator. Since we below will deal with what is essentially a two-state system, the natural projection operator follows closely from the spin-$1/2$ case.

The setup in this paper is based on the nonlinear interaction of light, through a Kerr medium. Thus, the founding description is given by a coupled set of nonlinear Schr\"odinger equations (NLSE). The NLSE has an enormous breadth of application, and can be found essentially whenever a wave system experiences slow modulations \cite{Sulem-Sulem, Scott}. Due to its structure, it naturally exhibits soliton formation, self-focusing, and wave collapse \cite{Sulem-Sulem}. Thus, it can be used for describing optical \cite{Ocean} and ocean rogue waves \cite{Shukla-etal}, the propagation of light through a nonlinear quantum vacuum \cite{Marklund-Shukla, Brodin-etal}, solitary waves in optical fibers \cite{Hasegawa}, atmospheric waves \cite{Stenflo}, and the formation of structures in plasma waves \cite{Plasma}. Thus, the NLSE allows for a very encompassing formulation which can be applied to a large variety of fields.

In this paper, we investigate the statistical dynamics of polarized light in a nonlinear refractive medium, starting from a two-state system described in terms of two coupled nonlinear Schr\"odinger equations (NLSEs). In particular, we obtain a generalized Wigner-Moyal type evolution equation for a scalar quasi-distribution function, where the polarization state is described by an extra independent variable. Here different polarization states corresponds to distinct positions on the Poincar\'e sphere for this variable. The conservation laws for our scalar distribution function are analyzed and the Vlasov limit is described. Furthermore, the modulational instability is studied for a fully coherent case, and a simple generalization to partially coherence is considered. Finally the applications and possible generalizations of our results are discussed.

\section{Fundamentals}

Two circularly polarized modes can be used to form a basis for a general polarization state. In this paper we consider propagation of two counter-rotating modes of right- and left-handed circularly polarized light in a guiding nonlinear Kerr medium, such as e.g. an optical fiber. The two modes are assumed to obey the following coupled system of nonlinear Schr\"odinger type evolution equations, c.f. \cite{TWWS, Agrawal},
\begin{align}
   & i\left(\frac{\partial\psi_+}{\partial z} + \frac{1}{v_{g+}}\frac{\partial\psi_+}{\partial x}\right) + \frac{\alpha}{2}\frac{\partial^2\psi_+}{\partial x^2}
    \label{eq:NLS+}   \\
    & = \frac{\beta_+}{2}\psi_+ + \frac{\kappa}{2}\psi_- + \gamma\left[(1-\nu)|\psi_+|^2+(1+\nu)|\psi_-|^2\right]\psi_+
    \nonumber
\end{align}
\begin{align}
    & i\left(\frac{\partial\psi_-}{\partial z} + \frac{1}{v_{g-}}\frac{\partial\psi_-}{\partial x}\right) + \frac{\alpha}{2}\frac{\partial^2\psi_-}{\partial x^2}
    \label{eq:NLS-}  \\
    & = \frac{\beta_-}{2}\psi_- + \frac{\kappa}{2}\psi_+ + \gamma\left[(1-\nu)|\psi_-|^2+(1+\nu)|\psi_+|^2\right]\psi_-
    \nonumber
\end{align}
where we have assumed the propagation direction to be along the $z$-axis with $x$ representing either the transverse coordinate or time depending on the situation under study. Correspondingly $\alpha$ represents either the diffraction or dispersion coefficient while $v_{g\pm}$ represents the group-velocity of the corresponding mode and $\gamma$ determines the strength of the nonlinearity. The system further includes both linear and nonlinear coupling with coupling coefficients $\kappa$ and $\nu$, respectively. For a low birefringence medium we have $\beta_+ = \beta_- = \beta_x + \beta_y$ and $\kappa = \beta_x - \beta_y$, with $\beta_x$ ($\beta_y$) being the linear propagation constant of the slow (fast) axis.

The system Eqs.(\ref{eq:NLS+}) and (\ref{eq:NLS-}) has been extensively studied in the context of nonlinear pulse propagation in optical fibers \cite{Agrawal}. It is a conservative system and can be shown to have the following Lagrangian

\begin{widetext}
\begin{align}
     & L = \sum_{n = +,-}\bigg\{\frac{i}{2}\left[\left(\frac{\partial\psi_n^*}{\partial z} + \frac{1}{v_{gn}}\frac{\partial\psi_n^*}{\partial x}\right)\psi_n - \psi_n^*\left(\frac{\partial\psi_n}{\partial z} + \frac{1}{v_{gn}}\frac{\partial\psi_n}{\partial x}\right)\right] + \frac{\alpha}{2}\left|\frac{\partial\psi_n}{\partial x}\right|^2 + \frac{\beta_n}{2}|\psi_n|^2 +\nonumber\\
     &+\frac{\gamma}{2}(1-\nu)|\psi_n|^4\bigg\} + \frac{\kappa}{2}\left(\psi_+^*\psi_- + \psi_-^*\psi_+\right) + \gamma(1+\nu)|\psi_+|^2|\psi_-|^2.
\end{align}
\end{widetext}

The aim of this paper is to introduce a scalar Wigner function and a concomitant evolution equation equivalent to the system Eqs.(\ref{eq:NLS+}) and (\ref{eq:NLS-}). To this end we define a correlation matrix $\Gamma_{mn}(x_1,x_2,z) = \langle\psi_m^*(x_1,z)\psi_n(x_2,z)\rangle$ and a Wigner matrix $W_{mn} = W_{mn}(x,p,z)$, viz.
\begin{equation}
    W_{mn}(x,p,z) = \frac{1}{2\pi}\int_{-\infty}^{\infty}\Gamma_{mn}(x+\xi/2,x-\xi/2,z)e^{i p\xi}d\xi,
    \label{eq:Wmn}
\end{equation}
where $x_1 = x + \xi/2$ and $x_2 = x - \xi/2$. The angle brackets $\langle\ldots\rangle$ denotes statistical averaging which will be used when considering extensions to partial coherence, see section \ref{sec:partial}. The Wigner transform method is frequently used in quantum theory and has also been successfully applied to the study of partially coherent light propagation in noninstantaneous nonlinear media. In the quasi-classical limit the formalism is further similar to the collision-less kinetic theory of plasma physics.

Given the correlation or Wigner matrix it is easy to construct the Stokes vector ${\bf S} = {\bf S}(x)$ which characterizes the polarization state by its mapping onto the Poincar\'e sphere. Introducing the extended set of Pauli matrices $\sigma_j$ ($\sigma_0 = I$) we have
\begin{equation}
    {\bf S}_j(x) = tr(\Gamma_{mn}(x_1,x_2)\sigma_j)\big|_{x_1=x_2=x}
\end{equation}
for the correlation matrix, with $tr(\ldots)$ denoting the trace which is equivalent to
\begin{equation}
    {\bf S}_j(x) = \int_{-\infty}^{\infty}tr(W_{mn}\sigma_j)~dp
\end{equation}
using the Wigner matrix. This agrees with the customary definition of the Stokes vector for circularly polarized light \cite{Jackson,BornWolf}. In component form we have
\begin{align}
    & {\bf S}_0(x) = \langle|\psi_+|^2\rangle(x) + \langle|\psi_-|^2\rangle(x) \\
    & {\bf S}_1(x) = 2\langle\psi_+\psi_-\cos(\delta)\rangle(x) \\
    & {\bf S}_2(x) = 2\langle\psi_+\psi_-\sin(\delta)\rangle(x) \\
    & {\bf S}_3(x) = \langle|\psi_+|^2\rangle(x) - \langle|\psi_-|^2\rangle(x),
\end{align}
with $\delta = \phi_- - \phi_+$ being the phase difference between the complex fields.

\section{Scalar Wigner Formalism}

To construct a scalar Wigner function we assume it to take the general form of a linear combination of the elements of the Wigner matrix. Specifically we exploit the analogy with spin-1/2 systems in the Wigner formalism of quantum theory \cite{ZMB} and define the scalar Wigner function as
\begin{equation}
    W(x,p,z,{\bf \hat{s}}) = \sum_{m,n}\frac{1}{2}\left[(1,{\bf \hat{s}})_j\cdot\sigma_j\right]_{mn}W_{mn}.
    \label{eq:scalarW}
\end{equation}
Note that this choice of representation is invertible but not unique. Another choice for a scalar Wigner function describing polarized light has previously been introduced by Luis in \cite{Luis}. The representation Eq.(\ref{eq:scalarW}) has the advantage of parameterizing the unit Poincar\'e sphere using the position vector ${\bf \hat{s}}$ and maps the circularly polarized states onto the poles of the sphere. It should be emphasized that we do not treat the polarization as a distribution but rather take the vector ${\bf \hat{s}}$ to determine the projection onto a particular state.

Using the definition of the Wigner transform Eq.(\ref{eq:Wmn}) it is easily found that the right- and left-handed circular intensities are given by
\begin{equation}
    \langle|\psi_+|^2\rangle = \int_{-\infty}^{\infty}W(x,p,z,{\bf \hat{z}})~dp
    \label{eq:I+}
\end{equation}
and
\begin{equation}
    \langle|\psi_-|^2\rangle = \int_{-\infty}^{\infty}W(x,p,z,-{\bf \hat{z}})~dp
    \label{eq:I-}
\end{equation}
while
\begin{equation}
    2\int_{-\infty}^{\infty}W(x,p,z,{\bf \hat{x}})~dp = \langle|\psi_+|^2\rangle + \langle|\psi_-|^2\rangle + 2\langle\psi_+\psi_-\cos(\delta)\rangle
\end{equation}
gives the intensity of the total field, including the beating term.

The scalar Wigner function $W(x,p,z,{\bf \hat{s}})$ can be seen as describing an instantaneous state of the system in an extended phase space. The evolution equation for this state is obtained by applying the Wigner transform to Eqs.(\ref{eq:NLS+}) and (\ref{eq:NLS-}), and then using the definition Eq.(\ref{eq:scalarW}). The particulars of similar derivations have previously been presented elsewhere \cite{Wigner,Dragoman}, thus we will only present the result of the calculation which is
\begin{widetext}
  \begin{align}
    & \frac{\partial W}{\partial z} + \frac{1}{2} \Bigg( \frac{1}{v_{g+}} + \frac{1}{v_{g-}} \Bigg) \frac{\partial W}{\partial x} + p \Bigg( \frac{1}{v_{g+}} - \frac{1}{v_{g-}} \Bigg) ({\bf\hat{s}} \times {\bf\hat{z}}) \cdot \nabla_{\bf \hat{s}} W + \frac{1}{2}  \Bigg( \frac{1}{v_{g+}} - \frac{1}{v_{g-}} \Bigg) \big[ {\bf \hat{z}}  \cdot \nabla_{\bf \hat{s}} + ({\bf \hat{s}} \cdot {\bf \hat{z}} ) (1-{\bf \hat{s}} \cdot \nabla_{\bf \hat{s}}) \big] \frac{\partial W}{\partial x} +  \nonumber \\
    &+\alpha p \frac{\partial W}{\partial x} - \left[{\bf \hat{s}}\times\left(\kappa{\bf \hat{x}} + \frac{\Delta\beta}{2}{\bf \hat{z}}\right)\right]\cdot\nabla_{\bf \hat{s}}W - 2\gamma\left(\langle|\psi_+|^2\rangle+\langle|\psi_-|^2\rangle\right)\sin\left(\frac{1}{2}\overleftarrow{\frac{\partial}{\partial x}}\overrightarrow{\frac{\partial}{\partial p}}\right)W +\nonumber\\
    & + 2\gamma\nu\left(\langle|\psi_+|^2\rangle-\langle|\psi_-|^2\rangle\right)\left[{\bf \hat{z}}\cdot\nabla_{{\bf \hat{s}}} + \left({\bf \hat{s}}\cdot{\bf \hat{z}}\right)\left(1-{\bf \hat{s}}\cdot\nabla_{{\bf \hat{s}}}\right)\right]\sin\left(\frac{1}{2}\overleftarrow{\frac{\partial}{\partial x}}\overrightarrow{\frac{\partial}{\partial p}}\right)W +\nonumber\\
    & + 2\gamma\nu\left(\langle|\psi_+|^2\rangle-\langle|\psi_-|^2\rangle\right)\left[\left({\bf \hat{s}}\times{\bf \hat{z}}\right)\cdot\nabla_{\bf \hat{s}}\right]\cos\left(\frac{1}{2}\overleftarrow{\frac{\partial}{\partial x}}\overrightarrow{\frac{\partial}{\partial p}}\right)W = 0,
    \label{eq:WMP}
\end{align}
\end{widetext}
where we have introduced the difference between the propagation constants, $\Delta\beta = \beta_+ - \beta_-$, and the intensities are provided by the relations Eqs.(\ref{eq:I+}) and (\ref{eq:I-}). Note that the trigonometric functions in Eq.(\ref{eq:WMP}) should be interpreted as operators defined by their respective series expansions, with the arrows indicating the direction of application of the derivatives.

Since the right- and left-handed circularly polarized intensities  do not figure independently in Eq.(\ref{eq:WMP}) but only in the combination of sum and difference intensities, it is convenient to write these in a shorthand form as
\begin{align}
    & I_0 = \langle|\psi_+|^2\rangle+\langle|\psi_-|^2\rangle
    \nonumber \\
    & = 2\int_{-\infty}^{\infty}W(x,p,z,{\bf 0})~dp = 2\int_{-\infty}^{\infty}\left(1-{\bf \hat{s}}\cdot\nabla_{{\bf \hat{s}}}\right)W~dp
\end{align}
and
\begin{equation}
    I_3 = \langle|\psi_+|^2\rangle-\langle|\psi_-|^2\rangle = 2\int_{-\infty}^{\infty}{\bf \hat{z}}\cdot\nabla_{{\bf \hat{s}}}W~dp.
\end{equation}
The evolution equation Eq.(\ref{eq:WMP}) together with the relations for the sum and difference intensities thus constitute a closed system of partial integro-differential equations. Although this system appears quite complicated, it can directly furnish projections onto a relevant evolution equation for arbitrary polarization states ${\bf \hat{s}}$. Additionally, the formalism highlights the role played by unequal group velocities $v_{g+} \neq v_{g_-}$ and unequal mode intensities $I_3 \neq 0$, which may not be obvious from Eqs.(\ref{eq:NLS+}) and (\ref{eq:NLS-}). The system can in general be shown to possess three invariants associated with the conservation of energy, of momentum and of the Hamiltonian, viz.
\begin{equation}
    E = \int_{-\infty}^{\infty}\int_{-\infty}^{\infty}\left(1-{\bf \hat{s}}\cdot\nabla_{{\bf \hat{s}}}\right)W~dp~dx = \textrm{const.}
\end{equation}
\begin{equation}
	M = \int_{-\infty}^{\infty}\int_{-\infty}^{\infty}p~\left(1-{\bf \hat{s}}\cdot\nabla_{{\bf \hat{s}}}\right)W~dp~dx = \textrm{const.}
\end{equation}
and
\begin{align}
    & H = \int_{-\infty}^{\infty}\Bigg[\alpha \int_{-\infty}^{\infty}p^2~(1 - {\bf \hat{s}}\cdot\nabla_{\bf \hat{s}})~W~dp
    \nonumber \\ &
    + 2\gamma\left(\int_{-\infty}^{\infty}(1 - {\bf \hat{s}}\cdot\nabla_{\bf \hat{s}})~W~dp\right)^2 - 2\gamma\nu\left(\int_{-\infty}^{\infty}{\bf \hat{z}}\cdot\nabla_{\bf \hat{s}}W~dp\right)^2 \nonumber\\
    & +\kappa\int_{-\infty}^{\infty}{\bf \hat{x}}\cdot\nabla_{\bf \hat{s}}W~dp + \frac{\Delta\beta}{2}\int_{-\infty}^{\infty}{\bf \hat{z}}\cdot\nabla_{\bf \hat{s}}W~dp \nonumber \\
    & + \bigg( \frac{1}{v_{g+}} - \frac{1}{v_{g-}} \bigg) \int_{-\infty}^{\infty} p~ {\bf \hat{z}} \cdot \nabla_{\bf {\hat{s}}} W~dp  \Bigg]dx = \textrm{const.}
\end{align}
These invariants are independent of the polarization direction and corresponds to the constant phase, x- and z-translation symmetries of the system Eqs.(\ref{eq:NLS+}) and (\ref{eq:NLS-}) by application of Noether's theorem. The evolution equation (\ref{eq:WMP}) is not known to have additional invariants, except in certain limits of the parameters, c.f. \cite{Hansson1}.

In the Vlasov limit the arguments of the trigonometric operators are small, i.e. $(\partial/\partial x),(\partial/\partial p)\ll 1$. Keeping only the lowest order contribution in an expansion, it turns out that $W$ is conserved along the orbits in the extended phase space.
 
As a consequence Eq.(\ref{eq:WMP}) can be written as
\begin{equation}
    \frac{dW}{d\zeta}=0
    \label{eq:orbit}
\end{equation}
with
\begin{equation}
    \frac{d}{d\zeta} \equiv \frac{\partial}{\partial z} + {\bf \hat{\chi}}\cdot\nabla_{\bf \hat{s}} + \frac{\partial f}{\partial p}\frac{\partial}{\partial x} - \frac{\partial f}{\partial x}\frac{\partial}{\partial p}
\end{equation}
where
\begin{align}
    & f = \frac{1}{2}\left(\frac{1}{v_{g+}}+\frac{1}{v_{g-}}\right)p + \frac{1}{2}\alpha p^2 + \gamma I_0 +\nonumber\\
    & +\left[\frac{1}{2}\left(\frac{1}{v_{g+}}-\frac{1}{v_{g-}}\right)p - \gamma\nu I_3\right]\left[{\bf \hat{z}}\cdot\nabla_{\bf \hat{s}} + ({\bf \hat{s}}\cdot{\bf \hat{z}})(1-{\bf \hat{s}}\cdot\nabla_{\bf \hat{s}})\right]
\end{align}
and
\begin{equation}
     {\bf \hat{\chi}}= -{\bf \hat{s}}\times\left\{\kappa{\bf \hat{x}} + \left[\frac{\Delta\beta}{2} - p\left(\frac{1}{v_{g+}}-\frac{1}{v_{g-}}\right) - 2\gamma\nu I_3\right]{\bf \hat{z}}\right\}.
\end{equation}
Since both $\partial^2 f/\partial x \partial p = 0$ and $\nabla_{\bf \hat{s}}\cdot {\bf \hat{\chi}} = 0$, we may alternatively write Eq.(\ref{eq:orbit}) as a continuity equation in the extended phase space, viz.
\begin{equation}
    \frac{\partial W}{\partial z} + \nabla_{\bf \hat{s}}\cdot({\bf \hat{\chi}} W) + \frac{\partial}{\partial x}\left(\frac{\partial f}{\partial p}W\right) - \frac{\partial}{\partial p}\left(\frac{\partial f}{\partial x}W\right) = 0.
\end{equation}
Which is an immediate consequence of the absence of photon creation and annihilation mechanisms.

\section{Modulational Instability}

It is well known that stationary monochromatic solutions of the system Eqs.(\ref{eq:NLS+}) and (\ref{eq:NLS-}) can experience modulation instability both in the anomalous ($\alpha\gamma < 0$) and normal ($\alpha\gamma > 0$) dispersion regimes. However, the stability analysis is complicated due to the general stationary continuous wave (CW) solution of Eqs.(\ref{eq:NLS+}) and (\ref{eq:NLS-}) being given in terms of Jacobian elliptic functions \cite{Agrawal}. Here, the Wigner transform method offers a convenient shorthand for obtaining a general dispersion relation, which illustrates many features of the polarization dependence, without the need to actually specify the background solution. For simplicity we consider the equal group velocity case $v_g \equiv v_{g+} = v_{g-}$ throughout.

To study the modulational instability associated with Eq.(\ref{eq:WMP}), c.f. \cite{LHAFSSH}, we linearize the evolution equations of the Wigner matrix elements using the ansatz $W_{mn} = W_{mn}^0 + \epsilon W_{mn}^1\exp\left[i(k x - \omega z)\right]$ and consider the first order perturbation of the stationary CW background solution $W_{mn}^0$ as $\epsilon \to 0$. To cast the resulting dispersion relation in a form suitable for the chosen Wigner distribution function we make the simplifying assumption of low birefringence, i.e. $\Delta\beta = 0$. After some calculations it is then found that the dispersion relation can be written as
\begin{widetext}
\begin{align}
    & 1 - 2\gamma\left(\overline{\frac{\Delta(1-{\bf \hat{s}}\cdot\nabla_{\bf \hat{s}})W^0}{\eta}} - \nu\overline{\frac{\Delta(1-{\bf \hat{s}}\cdot\nabla_{\bf \hat{s}})W^0}{\eta-4\kappa^2/\eta}}\right) -
     4\gamma^2\nu\left[\overline{\frac{\Delta(1-{\bf \hat{s}}\cdot\nabla_{\bf \hat{s}})W^0}{\eta}}~\overline{\frac{\Delta(1-{\bf \hat{s}}\cdot\nabla_{\bf \hat{s}})W^0}{\eta - 4\kappa^2/\eta}} - \overline{\frac{\Delta~{\bf \hat{z}}\cdot\nabla_{\bf \hat{s}}W^0}{\eta}}~\overline{\frac{\Delta~{\bf \hat{z}}\cdot\nabla_{\bf \hat{s}}W^0}{\eta - 4\kappa^2/\eta}}\right] +\nonumber\\
    & + 2\kappa\gamma\nu\overline{\frac{\Sigma~{\bf \hat{x}}\cdot\nabla_{\bf \hat{s}}W^0}{\eta^2-4\kappa^2}}\left(2\gamma\overline{\frac{\Delta(1-{\bf \hat{s}}\cdot\nabla_{\bf \hat{s}})W^0}{\eta}} - 1\right) - i4\kappa\gamma^2\nu\overline{\frac{\Delta~{\bf \hat{y}}\cdot\nabla_{\bf \hat{s}}W^0}{\eta^2-4\kappa^2}}~\overline{\frac{\Delta~{\bf \hat{z}}\cdot\nabla_{\bf \hat{s}}W^0}{\eta}} = 0
    \label{eq:general_drel}
\end{align}
\end{widetext}
where the overlines indicates integration with respect to $p$, $\Delta f(p) = f(p+k/2) - f(p-k/2)$, $\Sigma f(p) = f(p+k/2) + f(p-k/2)$ and $\eta = k\left(\frac{1}{v_g} + \alpha p\right) - \omega$.

Although, the general CW solution of the system Eqs.(\ref{eq:NLS+}) and (\ref{eq:NLS-}) is given in terms of elliptic functions it is also possible to obtain solutions in terms of elementary functions when the light is polarized along either the fast or slow axis \cite{Wabnitz}. The solution of Eq.(\ref{eq:WMP}) then takes the following simple form
\begin{equation}
    W = I_0\left(1 \mp s_x\right)\delta(p),
\end{equation}
which is seen to be independent of $s_y$ and $s_z$.

Another simplification is obtained for a purely incoherent interaction when the linear coupling term is neglected ($\kappa = 0$). The dispersion relation Eq.(\ref{eq:general_drel}) then reduces to
\begin{align}
    & 1 - 2\gamma(1-\nu)\overline{\frac{\Delta(1-{\bf \hat{s}}\cdot\nabla_{\bf \hat{s}})W^0}{\eta}}
    \label{eq:drel} \\ &
    - 4\gamma^2\nu\left[\left(\overline{\frac{\Delta(1-{\bf \hat{s}}\cdot\nabla_{\bf \hat{s}})W^0}{\eta}}\right)^2 - \left(\overline{\frac{\Delta~{\bf \hat{z}}\cdot\nabla_{\bf \hat{s}}W^0}{\eta}}\right)^2\right] = 0. \nonumber
\end{align}
Considering now a general coherent CW background solution of the form
\begin{align}
    & W^0 = \frac{1}{2}\big[(|\psi_+|^2 + |\psi_-|^2) + s_x(\psi_-^*\psi_+ + \psi_+^*\psi_-)
    \nonumber \\ &
    + i s_y(\psi_-^*\psi_+ - \psi_+^*\psi_-) + s_z(|\psi_+|^2 - |\psi_-|^2)\big]\delta(p).
\end{align}
It is seen that the modulational instability depends only on the total of the sum and difference intensities, regardless of the composition of the fields, since
\begin{equation}
    (1 - {\bf \hat{s}}\cdot\nabla_{\bf \hat{s}})W^0 = \frac{1}{2}I_0\delta(p), \qquad {\bf \hat{z}}\cdot\nabla_{\bf \hat{s}}W^0 = \frac{1}{2}I_3\delta(p),
    \label{eq:bdist}
\end{equation}
where both $I_0$ and $I_3$ are constant since the modes energies are now individually conserved.

In this case the dispersion relation Eq.(\ref{eq:drel}) can be evaluated explicitly with the result
\begin{equation}
    \omega = \frac{k}{v_g} \pm \frac{|\alpha k|}{2}\sqrt{k^2+\frac{2\gamma}{\alpha}(1-\nu)I_0 \pm \frac{2\gamma}{\alpha}\sqrt{(1+\nu)^2I_0^2-4\nu I_3^2}}.
    \label{eq:disp_rel}
\end{equation}
Eq.(\ref{eq:disp_rel}) clearly shown that the modulational instability can occur also in the normal dispersion regime due to the nonlinear cross phase modulation, since for $0 \leq I_3\leq I_0$ we have that $(1+\nu)^2I_0^2\geq(1+\nu)^2 I_0^2-4\nu I_3^2\geq(1-\nu)^2I_0^2$. The instability growth rate is given by
\begin{equation}
    \Omega = \frac{|\alpha k|}{2}\sqrt{\frac{2\gamma}{\alpha}\left[\sqrt{(1+\nu)^2I_0^2-4\nu I_3^2}-(1-\nu)I_0\right]-k^2},
\end{equation}
c.f. the corresponding expression for the partially coherent case in \cite{LHAFSSH}. In the anomalous dispersion regime the instability growth rate analogously becomes
\begin{equation}
    \Omega = \frac{|\alpha k|}{2}\sqrt{\frac{2\gamma}{\alpha}\left[(1-\nu)I_0 + \sqrt{(1+\nu)^2I_0^2 - 4\nu I_3^2}\right] - k^2}.
\end{equation}
In either case the maximum region of instability and the maximum growth rate are obtained when the mode intensities are equal, i.e. $I_3 = 0$.

\section{Partial Coherence}
\label{sec:partial}

The above considerations can easily be extended to include the effects of partial coherence by allowing the fields to have a stochastically varying phase with statistical properties described by the mutual correlations between the fields \cite{Hansson2}. The partial coherence manifest itself as an additional width in $p$-space of the Wigner distribution function, beyond the intrinsic spectral width of the coherent fields. In general, the coherence properties of each independent element of the hermitian correlation matrix needs to be specified separately.

When considering modulational instability, we assume, for analytical simplicity, that the statistical properties of the CW background solution for the two counter-rotating fields are identical, as would be the case if e.g.~they were generated by the same source. We further assume that their coherence properties can be described by a Lorentzian background distribution, viz.
\begin{equation}
    (1 - {\bf \hat{s}}\cdot\nabla_{\bf \hat{s}})W^0 = \frac{I_0}{2\pi}\frac{p_0}{p^2+p_0^2}, \qquad {\bf \hat{z}}\cdot\nabla_{\bf \hat{s}}W^0 = \frac{I_3}{2\pi}\frac{p_0}{p^2+p_0^2},
\end{equation}
where $p_0^{-1}$ is a parameter which can be identified with the correlation length. Note that in the coherent limit of $p_0 \to 0$, the Lorentzian approaches a delta function and Eq.(\ref{eq:bdist}) is recovered as necessary.

For the case of $\kappa = 0$, the dispersion relation Eq.(\ref{eq:drel}) may again be evaluated in a closed form with the result
\begin{align}
    & \frac{1}{|\alpha k|}\left(\omega-\frac{k}{v_g}\right)
    \label{eq:disp_rel2} \\ &
     = \pm \frac{1}{2}\sqrt{k^2+\frac{2\gamma}{\alpha}(1-\nu)I_0 \pm \frac{2\gamma}{\alpha}\sqrt{(1+\nu)^2I_0^2-4\nu I_3^2}} - i p_0, \nonumber
\end{align}
c.f. Eq.(\ref{eq:disp_rel}) and \cite{LHAFSSH}. The partial coherence thus acts to suppress the instability growth rate in a similar fashion to the Landau damping of electron plasma waves. Indeed, for a sufficiently large degree of incoherence the instability may be completely quenched.

\section{Conclusions}

We have proposed a scalar Wigner distribution function for describing polarized light propagating in a nonlinear Kerr medium. By employing a circular polarization basis the polarization state is included in a manner similar to the spin state in the quantum kinetic theory of spin-1/2 particles. The approach has been applied to optical propagation in a Kerr medium governed by a coupled system of nonlinear Schr\"odinger equations and an extended phase space evolution equation has been derived along with invariant quantities. We have further considered the problem of modulational instability and derived a dispersion relation valid for general background distributions, which in particular is shown to agree with the expected result in the limit of low birefringence and no linear coupling. Finally the extension to partially coherent fields and its effect on the modulational instability has been considered.

Future development of the above results could be in the direction of a 3D theory, as well as allowing for structurally different types of interactions (i.e., leaving the limitations of Kerr media). Such developments could prove of interest in e.g. magnetization studies through short laser pulses, where the statistical properties of the incoming light, and its interaction with the matter target, can give rise to interesting new dynamics of spin structures. Thus, the coupling between  different degrees of freedom in light and matter would be able to capture a large set of fast dynamics. Moreover, such generalizations could also include a more generic description of the invariants of the field. Such a development would allow for a less than ad hoc formulation of the spin and angular momentum properties of light in terms of a scalar distribution function. It is well known that the angular moment of light can readily affect the dynamics of matter structures; it would be of interest to see how such a formulation could highlight the interaction between partial coherence, spin/angular momentum, and matter dynamics induced by light.

\acknowledgements
This research was supported by the European Research Council, contract 204059-QPQV, and the Swedish Research Council, contract 2010-3727.

\end{document}